\documentclass[12 pt]{article}

\usepackage{amsmath}
\usepackage{amssymb}
\usepackage{graphicx}
\usepackage{subfigure}

\begin{document}

\ \vskip 0.5 in
\begin{center}
{ \large {\bf Quantum Measurement and Quantum Gravity\footnote {Expanded version of 
arXiv:0705.2357 [gr-qc]. Based on talks given at: [FTAGVI, HRI, Allahabad, 13-18 Nov. 2007]; [Himalayan Relativity Dialog, Mirik, April, 2007]; [Workshop Session on Quantum Gravity, IAGRG, Delhi, Feb. 2007];
[Parmenides Workshop: The present - perspectives from physics and philosophy, Wildbad Kreuth, Germany, October, 2006]; [Workshop Session on Quantum Gravity, ICGC, Pune, Dec. 2007]. }  }}\\
{ {\bf - Many-Worlds or Collapse of the Wave-Function?  - }}

\bigskip

\bigskip

{{\large
{\bf T. P. Singh\footnote{e-mail address: tpsingh@tifr.res.in} }}}

\medskip

{\it Tata Institute of Fundamental Research,}\\
{\it Homi Bhabha Road, Mumbai 400 005, India.}
\vskip 0.5cm
\end{center}

\smallskip

\begin{abstract}

\noindent At present, there are two possible, and equally plausible, explanations for the physics of quantum measurement. The first explanation, known as the many-worlds interpretation, does not require any modification of quantum mechanics, and asserts that at the time of measurement the Universe splits into many branches, one branch for every possible alternative. The various branches do not interfere with each other because of decoherence, thus providing a picture broadly consistent with the observed Universe. The second explanation, which requires quantum mechanics to be modified from its presently known form, is that at the time of measurement the wave-function collapses into one of the possible alternatives. The two explanations are mutually exclusive, and up until now, no theoretical reasoning has been put forward to choose one explanation over the other. In this article, we provide an argument which implies that the collapse interpretation is favored over the many-worlds interpretation. Our starting point is the assertion (which we justify) that there ought to exist a reformulation of quantum mechanics which does not refer to a classical spacetime manifold. The need for such a reformulation implies that quantum theory becomes non-linear on the Planck mass/energy scale. Standard linear quantum mechanics is an approximation to this non-linear theory, valid at energy scales much smaller than the Planck scale. Using ideas based on noncommutative differential geometry, we develop such a reformulation and derive a non-linear Schr\"{o}dinger equation, which can explain collapse of the wave-function. We also obtain an expression for the lifetime of a quantum superposition. We suggest ideas for an experimental test of this model.

\end{abstract}

\newpage

\tableofcontents

\vskip 1.0 in

\noindent  {\bf {\textit{" I would like to suggest that it is possible that quantum mechanics fails for large distances and large objects. Now, mind you, I do not say that quantum mechanics does fail at large distances, I only say that it is not inconsistent with what we do know. If this failure of quantum mechanics is connected with gravity, we might speculatively expect this to happen for masses such that
$GM^{2}/\hbar c=1$, of $M$ near $10^{-5}$ grams, which corresponds to some $10^{18}$ particles  "}}}

\bigskip

\rightline{- Feynman (1957)}

\newpage
\section{The Quantum Measurement Problem}

Suppose that in a quantum measurement one measures an observable $\hat{O}$ of a quantum system, and suppose that prior to the measurement the system is in a state $|\psi>$ which can be expanded in a basis of orthonormal eigenstates $|\psi_n>$ of the observable $\hat{O}$ as
\begin{equation}
|\psi> = \sum_n a_n |\psi_n>.
\label{exp}
\end{equation}
It is then known from experiment that, if the selected states of the measuring apparatus are in one-to-one correspondence with the basis $|\psi_n>$, then after the measurement the  quantum system is found to be in one of the eigenstates, say $|\psi_n>$. Repeated measurements on identical copies of the quantum system show that the system is found to be in one or the other eigenstates $|\psi_n>$, with the probability to be found in state $|\psi_n>$ being given by $|a_n|^2$ (the Born probability rule).

The transition $|\psi>\rightarrow|\psi_n>$ that takes place during a quantum measurement cannot be described by the Schr\"{o}dinger equation. This is of course because the Schr\"{o}dinger equation is linear, and Schr\"{o}dinger evolution will preserve the superposition expressed in Eqn. (\ref{exp}). The observed transition $|\psi>\rightarrow|\psi_n>$, on the other hand, breaks linear superposition. The quantum measurement problem can be stated as follows: what is the correct physical description of this measurement process and of the observed result? This description should explain why the transition takes place in the first place, and why it obeys the Born probability rule. There are two possible explanations, which we elaborate on below.

\subsection{First Explanation : The Many-Worlds Interpretation}
According to the many-worlds interpretation of quantum mechanics, (originally due to Everett 
\cite{Everett}), despite appearances, the transition $|\psi>\rightarrow|\psi_n>$ in fact does not take place at all during a quantum measurement, and superposition continues to be preserved after the measurement has taken place. Rather, it is assumed that during a quantum measurement the Universe splits into many branches, with a particular outcome, say $|\psi_n>$, being realized in our branch of the Universe. Here, the term Universe is meant to refer also to the measuring apparatus, and the observer as well. It has also been asserted that the many-worlds interpretation is consistent with the Born probability rule \cite{mwiborn}. (For another discussion on the probability rule in the
many-worlds picture see \cite{zee}.)
 A recent account of the many-worlds interpretation, including recent developments in its understanding, can be found in \cite{mwirev}. 
  
The different branches of the Universe do not interfere with each other because of the phenomenon of
decoherence. The process of decoherence, which has been experimentally observed \cite{decohexpt}, 
destroys the interference amongst various alternatives in the quantum state of a macroscopic system 
consisting of a superposition of various alternatives \cite{kiefer}. Thus in the many-worlds picture the various branches of the Universe continue to remain superposed, as required by Schr\"{o}dinger evolution,
but do not interfere with each other, as a consequence of decoherence. In this picture, quantum mechanics does not have to be modified in order to explain a quantum measurement. This has been called `economy of assumptions, and extravagance of Universes'.

The many-worlds interpretation may appear counter-intuitive, but it is difficult to find logical
inconsistencies in the interpretation. The picture is in fact attractive because it can explain quantum measurement without having to change the laws of quantum mechanics. Its shortcoming perhaps is that
it does not appear to be experimentally falsifiable (unless experimental proof can be found for the `collapse' interpretation discussed next). Also, if the Universe splits into many branches, it is not clear how the obscure issue of `splitting of the consciousness of an observer into many branches'
is to be understood. Neither of these shortcomings however can by themselves rule out the possibility
that the many-worlds interpretation could be the correct explanation of quantum measurement.

\subsection{Second Explanation: Collapse of the Wave-function} 
The second explanation is that the transition $|\psi>\rightarrow|\psi_n>$ does indeed take place, and the Universe does not split into many branches. It is assumed that there is only one branch to the Universe, the one that we directly observe, and live in. The two explanations (many-worlds and collapse) are clearly mutually exclusive : one, and only one, out of the two explanations must be correct.

In order to explain quantum measurement by invoking collapse of the wave-function, quantum mechanics
and the Schr\"{o}dinger equation must be modified. Quantum mechanics as we presently know it can only be an approximation to a more general theory, with the more general theory having the capacity to explain
wave-function collapse. This has been called `economy of Universes, and extravagance of assumptions'.

For instance, it may be possible to explain quantum measurement by generalizing the Schr\"{o}dinger equation to a non-linear Schr\"{o}dinger equation. The non-linearity is assumed to become important in the measurement domain, but is negligible in the microscopic domain. In principle, the presence of the
non-linearity can result in breakdown of superposition, driving the quantum system to one particular alternative, in a manner consistent with the Born probability rule. This particular approach to collapse
of the wave-function will be the focus of the present paper. Other models of collapse are briefly reviewed in Section 4. Objections against a non-linear quantum mechanics (such as superluminality) are briefly addressed in the Discussion section.

It is only fair to say that as of now, there is no universally accepted theory for collapse of the wave-function, supported by experiment. Neither is there any experimental evidence that quantum mechanics has to be modified from its present form. Needless to say, this situation could change in the future.
 
\subsection{Goal of the Present Paper}
Up until now, there has been no experimental or theoretical motivation to favor the many-worlds interpretation of quantum measurement over the collapse interpretation, or vice-versa. Critics of the
many-worlds interpretation could say its unfalsifiable, while adherents of this interpretation consider it an advantage that it requires no changes in the existing formulation of quantum theory. Proponents of
the collapse model find it unphysical that unobservable parallel Universes are invoked so as to protect the prevailing structure of quantum theory, whereas critics label the collapse models as \textit{ad hoc}; having been invented for the sole purpose of explaining quantum measurement, and not embedded in a broader theoretical framework.

The purpose of the present paper is to argue that there are additional theoretical reasons, hitherto unemphasized, which suggest that the collapse explanation is favored over the many-worlds interpretation. Our starting point, which we elaborate on in detail in Section 3, is that the notion
of time which is used to describe time evolution in quantum theory is a classical notion. There
ought to exist a reformulation of quantum mechanics which does not make a reference to this external
classical notion of time. Furthermore, under appropriate circumstances, i.e. as and when one chooses to make reference to (a possibly available) external classical time, this reformulation should become equivalent to standard quantum mechanics.

The central thesis of our paper is that the requirement that there be such a reformulation of quantum mechanics leads to the conclusion that standard quantum theory is a limiting case of a more general, non-linear, quantum theory. The non-linearity becomes important in the measurement domain, and could cause the collapse of the wave-function during a quantum measurement. This is why we say that the collapse based explanation is favored over the many-worlds interpretation. We arrive at this conclusion, not in an \textit{ad hoc} fashion, but by addressing an entirely different incompleteness in the existing formulation of quantum theory, namely, its undesirable reference to an external classical time. Thus our conclusion is that removing the notion of classical time from quantum mechanics has an important and significant byproduct - one may be able to explain quantum measurement as a consequence of wave-function collapse.
The requirement that the aforementioned reformulation of quantum mechanics should exist is unavoidable;
and non-linearity in quantum mechanics is its inevitable consequence. The many-worlds interpretation is
thus disfavored, not because it might appear to be counter-intuitive and discomforting, but because of 
compelling theoretical reasoning which extends outside and beyond the current formulation of quantum mechanics. 

In Section 3 we will propose a tentative reformulation of quantum mechanics, borrowing ideas from noncommutative geometry. We will arrive at a non-linear Schr\"{o}dinger equation which generalizes the standard linear Schr\"{o}dinger equation, and which can explain at least some significant aspects of the process of quantum measurement.

Before we present this reformulation, we will review in the next Section an illustrative toy-model for collapse induced by non-linearity, which is due to Grigorenko \cite{Grigorenko}. This model will be of help to us in understanding the relation between wave-function collapse and the non-linear equation we arrive at in Section 3. 

[A few other significant interpretations of quantum mechanics, not discussed in the present paper, have also been proposed in the literature. These include the works of Bohm \cite{Bohm} (and its
generalization to non-local hidden variable theories), and those of  Gell-Mann and Hartle \cite{GH}, and Omnes \cite{Omnes}. These works present a different formulation of standard quantum
mechanics, without implying that the experimental predictions of the new formulation differ from
that of the standard theory.]

\section{A Toy Model for Non-linear Quantum Mechanics and Collapse of the Wave-function}

\subsection{Introduction}

We propose a non-linear Schr\"{o}dinger equation which during a quantum measurement can dynamically induce the transition $|\psi>\rightarrow|\psi_n>$ with a probability $|a_n|^{2}$, when the initial state of the quantum system is given by Eqn. (\ref{exp}). Since such a  non-linear equation must not violate what is known about quantum mechanics from experiments (including stringent bounds on non-linearity in the microscopic domain), it must satisfy several conditions, which we enumerate below:

\begin{itemize}

\item The non-linearity must become significant only during the onset of a quantum measurement, and not before that.

\item The non-linear equation must reduce to the standard linear Schr\"{o}dinger equation when applied to the special case of microscopic systems.

\item The non-linear equation must reduce to the standard equation of motion of classical mechanics when applied to the special case of macroscopic systems.

\item The Hamiltonian must have a non-Hermitean part, which is also non-linear, so that the initial superposition of states can decay into one of the alternatives. However, the action of the 
full Hamiltonian on the states must be norm-preserving.  

\item Since in this scenario the outcome of a quantum measurement is deterministic but random,
the non-linear equation must contain one or more random variables. One or the other outcome of 
a measurement is realized, in consistency with the Born rule, depending on the relative values of the
random variables.

\item The predictions of the non-linear equation must be experimentally testable, and must not contradict
the experimentally verified features of standard quantum mechanics.

\item The non-linear equation must not be \textit{ad hoc}, but must instead be a consequence of `some other requirement'. In other words, there must be good theoretical reasons, independent of quantum measurement, such as those outlined in Section 1.3, for such a non-linear generalization of quantum mechanics. 
 
\end{itemize}

In this Section, we will work with a non-linear Schr\"{o}dinger equation which belongs to the following general class of norm-preserving non-linear Schr\"{o}dinger equations:
\begin{equation}
i\hbar d|\psi>/dt = H|\psi> + (1-P_\psi)U|\psi>.
\label{gri}
\end{equation}
Here, $H$ is the Hermitian part of the Hamiltonian, as in standard quantum mechanics. 
$(1-P_\psi)U$ is the
non-Hermitian part, $P_\psi=|\psi><\psi|$ is the projection operator, and $U$
is an arbitrary  nonlinear operator. This equation has been discussed by \cite{Gisin} and
by \cite{Grigorenko}. We will make the crucial assumption (to be justified later, in Section 3) that the non-Hermitean part of the Hamiltonian becomes significant only when the mass of the system becomes comparable to or larger than Planck mass, $m_{Pl}=(\hbar c/G)^{1/2}\sim 10^{-5}$ grams. 

We use the term `initial system' to refer to the quantum system ${\cal Q}$
on which a measurement is to be made by a classical apparatus ${\cal A}$, 
and the term `final system' to refer jointly to 
${\cal Q}$ and ${\cal A}$ after the initial system has interacted with 
${\cal A}$. A quantum measurement will be thought of as an increase in the
mass (equivalently, number of degrees of freedom) of the system, from     
the initial value $m_{\cal Q}\ll m_{Pl}$ to the final value 
$m_{\cal Q} + m_{\cal A} \gg m_{Pl}$. Clearly then, the non-Hermitian part of the Hamiltonian will play a crucial role in the transition from the initial system to the final system. 

We assume that ${\cal A}$ measures an observable ${\hat O}$ of ${\cal Q}$, 
having a complete set of eigenstates $|\psi_n>$. Let the quantum state of the
initial system be given as $|\psi>=\Sigma_n\ a_n|\psi_n>$. 
The onset of 
measurement corresponds to mapping the state $|\psi>$ to 
the entangled state $|\psi>_F$ of the final system as
\begin{equation}
|\psi>\rightarrow |\psi>_F\ \equiv  \sum_n a_n|\psi>_{Fn}= \sum_n\ a_n|\psi_n>|A_n>
\label{map0}
\end{equation}
where $|A_n>$ is the state the measuring apparatus would result in, had the 
initial  system been in the state $|\psi_n>$. During a quantum measurement the non-Hermitian part of the Hamiltonian
dominates over the Hermitian part, and governs the evolution
of the state $|\psi>_F$.

\subsection{The Toy Model}
As a useful and illustrative toy-model, we will consider a special case of the non-linear equation 
(\ref{gri}) with the operator $U$ given by
\begin{equation}
U=i\gamma\sum_n q_n |\psi>_{Fn}<\psi|_{Fn},   \qquad \qquad H=0,   
\label{you}
\end{equation}
where the $q_n$ are random real, positive constants \cite{Grigorenko}. $\gamma$ is a constant which will be assumed to be zero before the onset of measurement, and non-zero during the measurement. This is consistent with the assumption, made above, that the non-Hermitean part of the Hamiltonian becomes significant only when the mass of the system becomes comparable to or larger than Planck mass. Thus by system here we mean the `final system', which jointly refers to the measuring apparatus ${\cal A}$ and the initial quantum system ${\cal Q}$. The state $|\psi>_{Fn}$ has been defined in (\ref{map0}) above. The remaining Hermitean part of the Hamiltonian has been set to zero for simplicity; because here we want to demonstrate how the 
non-Hermitean part is responsible for the decay of the superpositions initially present in the quantum system $Q$. Including the Hermitean part will not prevent the breakdown of superposition - it will only
make the analysis more complicated.  

[We are using Grigorenko's model to illustrate collapse induced by non-linearity, because of its
simplicity. However, the essential idea has been put forward much earlier, probably first by
Bohm and Bub \cite{BohmBub}, who suggested a non-linear Schr\"{o}dinger equation to explain
measurement, using hidden variables as random variables. More pertinent to our context is a 1976 paper  
by Pearle \cite{Pearle} who proposed a non-linear Schr\"{o}dinger equation with phases of states as random variables, to explain measurement. This insightful and highly readable paper already put
forth, in spirit, the non-linear mechanism proposed by us, except that a fundamental origin for
the non-linearity was not suggested there. We find it curious that subsequent focus shifted mainly to
other dynamical reduction models (not involving the phase as a random variable) and Pearle's original suggestion was largely forgotten. In a sense, our present work revives Pearle's idea, bringing it in as a byproduct of considerations originating in quantum gravity.]

Let us now analyze Eqn. (\ref{gri}), with the understanding that the state $|\psi>$ is here $|\psi>_F$.
Substituting the form of the Hamiltonian given in (\ref{you}) into the non-linear Schr\"{o}dinger equation (\ref{gri}) gives, after using the expansion for $|\psi>_{Fn}$ given in (\ref{map0}),
\begin{equation}
da_n/dt = \gamma a_n (q_n -L)
\end{equation}
where $L=\Sigma_n q_n |a_n|^2$. Hence we can write
\begin{equation}
\frac{d}{\gamma dt}\left( \ln \frac{a_i}{a_j}\right) = q_i - q_j.
\end{equation}

Since the evolution is norm-preserving, 
it is easily inferred that the system evolves to the state with the largest value of $q_n$. In a 
repeated measurement, different outcomes can be achieved by different sets of values of the random variables $q_n$. For instance, if the $q_n$ each lie in the range $[0,\infty]$ and have a probability distribution
\begin{equation}
\omega(q_n) = |<\psi(t_0)|\psi_n>|^2 \exp(|<\psi(t_0)|\psi_n>|^2 q_n) 
\end{equation}
then the probability that the system evolves to the state $|\psi_n>$ can be shown to be
$|<\psi(t_0)|\psi_n>|^2$, as required by experiments \cite{Grigorenko}.

More details and interesting features of this and related models can be found in Grigorenko's paper. For instance, the number of random variables can be less than the number of system states. Furthermore, an attractive candidate for a random variable is the phase of the initial quantum state $\psi(t_0)$.
As explained by Grigorenko, as the initial phase varies randomly and uniformly in the range $[0,2\pi]$,
it is possible to have different outcomes in a repeated measurement, depending on the value of the initial phase, and in consistency with the Born probability rule. Grigorenko also discusses how to
overcome the problem of superluminality in his model.

The implication of this model is that evolution during a quantum measurement is deterministic. 
Probabilities enter the picture because of the presence of one or more random variables in a non-linear
Schr\"{o}dinger equation with a non-Hermitean part. Non-linearity breaks superposition. The non-Hermitean part causes decay/growth of different components of the quantum state.  The presence of the random variables ensures that in a
repeated measurement, different outcomes are realized, and the Born probability rule can be recovered
by associating suitable probability distributions with the random variables.

Nevertheless, this is a toy model, and the form of the Hamiltonian is decidedly \textit{ad hoc}. In Section 3 we will show that a Hamiltonian with the same features as in this toy model arises from addressing a 
fundamental incompleteness of quantum mechanics - the presence of an external classical time in the theory. We will hence be able to provide a collapse based explanation of quantum measurement, for reasons having to do with quantum gravity. We also note that the toy model has a limitation that the form of
the non-linear operator can only be written down {\it after} choosing a basis. This limitation
will not arise for the non-linear equation derived in Section 3.

\subsection{The Doebner-Goldin Equation}
In this brief digression we point out an important non-linear Schr\"{o}dinger equation which belongs to the general class (\ref{gri}). This is the Doebner-Goldin equation \cite{dg}
\begin{equation}
i\hbar \frac{\partial\psi}{\partial t} = -\frac{\hbar^{2}}{2m}\nabla^{2}\psi + iD\hbar\left(
\nabla^{2}\psi + \frac{|\nabla\psi|^{2}}{|\psi|^{2}}\psi\right).
\label{dg}
\end{equation} 
Here, $D$ is a real constant. The origin of the D-G equation has been discussed in detail in \cite{dg}; its generalizations are discussed by Goldin \cite{Goldin} and also in \cite{singh}. 

With some further assumptions, the D-G equation can be used to explain the collapse of the 
wave-function \cite{tpe}, in the manner of the toy-model described above. The non-linear Schr\"{o}dinger equation that we arrive at in the next Section is very similar to the above D-G equation.

\section{Quantum Gravity suggests that Quantum Mechanics is Non-linear} 

\subsection{Outline of the approach}

We outline below the key steps in the development of our intended non-linear theory :

\begin{itemize}

\item There ought to exist a reformulation of quantum mechanics 
which does not refer to a classical spacetime manifold. This provides a new path to quantum
gravity.

\item It then follows as a consequence that quantum theory as we know it is a limiting case of a 
non-linear quantum theory.

\item We propose the desired reformulation of quantum mechanics using ideas from noncommutative 
differential geometry.

\item This has implications for the quantum measurement problem : we arrive at a non-linear Schr\"{o}dinger equation, with the non-linearity becoming significant in the vicinity of the Planck-mass scale.

\end{itemize}

A detailed discussion of the approach can be found in \cite{singh}.

\subsection{Why quantum mechanics without classical spacetime?}
The concept of time that appears in quantum mechanics is a classical concept.
It is part of a classical spacetime manifold. The overlying metric on this spacetime manifold
is produced by classical matter fields. In our present Universe we take the presence of such classical matter fields as given. 
But the Universe could in principle be in a state in which there are no classical matter fields.

If only quantum matter fields are present in the Universe, then the metric produced by them will undergo quantum fluctuations. If the metric is undergoing quantum fluctuations, then one cannot assign
physical significance to the underlying classical spacetime manifold.
This is the essence of the Einstein hole argument \cite{Einstein}, \cite{Christian}.
Hence it is necessary to have a fundamental reformulation of quantum mechanics which does not refer
to a classical spacetime manifold.

\bigskip

\noindent [{\bf  The Einstein hole argument :} 
Consider a spactime manifold ${\cal M}$ having matter
fields, except in a hole $H$ inside the manifold, which is devoid of matter fields. The only field present in the hole is the gravitational field, described by the metric $g_{\mu\nu}$. Consider an active
diffeomorphism $\phi$ on {\cal M}, which is by definition identity outside the hole, and on its boundary, but different from identity inside the hole. Clearly, the stress-tensor $T_{\mu\nu}$ remains unchanged under this diffeomorphism, all over the spacetime. However, inside the hole, the metric changes under the diffeomorphism, say from
$g_{\mu\nu}(p)$ at the point $p$ to $\phi * g_{\mu\nu}(q)$ at the point $q$, where $q$ is the point to which the point $p$ is mapped, by the active diffeomorphism. Since $T_{\mu\nu}$ has not changed, it
is natural to expect that the physical gravitational field has not changed either. This is achieved by
appealing to general covariance, namely that the metrics $g_{\mu\nu}(p)$ and $\phi * g_{\mu\nu}(q)$
describe the same physical gravitational field.

However, Einstein realized that general covariance comes at a price. The fields $g_{\mu\nu}(p)$ and 
$\phi * g_{\mu\nu}(q)$ can be regarded as physically identical only if the points $p$ and $q$ are also
regarded as physically identical! It is thus a consequence of general covariance that points on a spacetime manifold cease to have any physical meaning as separate, distinct points of a spacetime. The only way to restore a physical attribute to points of the spacetime manifold (i.e. consider them as events) is through the presence, on the manifold, of a specific dynamically determined metric tensor field - the metric tensor field serves to provide a label to the points.

In our present context, we hence see that if the metric tensor is undergoing quantum fluctuations, one
can no longer regard the underlying classical spacetime manifold as having a physically meaningful 
status \cite{carlip} . ]

As and when the Universe is in a state in which it is dominated by classical matter fields, a classical
spacetime manifold, and a classical metric become available. Quantum mechanics could then be equivalently described in two ways: (i) either in the proposed reformulation, which continues to avoid making
reference to the time which is now externally available, or, as we always do (ii) from the point of view of an observer in the classical Universe, as standard quantum mechanics wherein evolution is described
with respect to an external time. The proposed reformulation should become equivalent to the standard
formulation when an external time becomes available.

In such a reformulation there is no classical spacetime; however, we can envisage the concept of a `quantum spacetime' and a `quantum gravitational field' which is produced by quantum matter fields. The `quantum gravitational field', like its classical counterpart, is assumed to act as a source for itself. This makes the quantum theory of gravity a non-linear theory - this feature is central to the thesis of
this paper. Such a theory is completely different from the Wheeler-DeWitt equation, wherein
the quantum theory of gravity is linear by construction. 

A non-linear quantum gravity might appear counterintuitive because we are generally conditioned to building a quantum theory by `quantizing' a classical theory. These rules of quantization are linear by definition, and further, they assume an external classical time as given. However, when such an external classical time is no longer available, we do not have such rules of quantization ready at hand. We are compelled to adopt a top-down approach, and then we see that it is very natural that `quantum gravity' `quantum gravitates', in precisely the same sense in which classical gravity acts as a source for 
itself -  thus leading to a non-linear quantum gravity. What this means, for instance, is that if we were to write an analog of the Wheeler-DeWitt equation for the non-linear case, the Hamiltonian of the theory would depend on the quantum state.

As has been explained in detail in \cite{singh}, this non-linearity in the quantum gravity theory becomes significant only at the Planck mass/energy scale. At much lower energy scales, the theory is linear,
to an excellent approximation.
This has an analogy with classical general relativity - the classical theory is non-linear only in the strong-field regime, but linear in the weak field Newtonian regime.

Next, we consider what the equation of motion of a quantum field or a quantum mechanical particle  is, in such a `quantum spacetime'. Again, as shown in detail in \cite{singh}, if the mass of the particle is comparable to Planck mass, its quantum dynamics is influenced by its own quantum gravitational field, and the equation of motion is non-linear. From the point of view of an external spacetime (as and when the latter becomes available) this equation of motion (in the non-relativistic limit) will appear to be a non-linear Schr\"{o}dinger equation. When the particle's mass is much smaller than Planck mass, the
non-linear equation will reduce to the standard linear Schr\"{o}dinger equation.  

\subsection{A Reformulation Based on Noncommutative Differential Geometry}
We suggest the concept of a noncommuting coordinate system, which `covers' a noncommutative
manifold, wherein commutation relations between coordinates are 
introduced on physical grounds. Commutation relations amongst momenta
must also be introduced. 

Our proposal is that basic laws are invariant under general coordinate 
transformations of noncommuting coordinates.This generalizes the standard concept of general covariance
to noncommuting coordinates.  This formulation should satisfy two important properties:

\begin{itemize}

\item Firstly, in the limit in which the system becomes macroscopic,
the noncommutative spacetime should be indistinguishable from  ordinary
commutative spacetime, and the dynamics should reduce to classical dynamics. 

\item Secondly, if a dominant part of the system becomes macroscopic and 
classical, and a sub-dominant part remains quantum (as our Universe is) then 
seen from the viewpoint of the dominant part, the quantum dynamics of the 
sub-dominant part should be the same as the standard quantum dynamics known to us. 

\end{itemize}

Our overall proposal for the reformulation of quantum mechanics in the language of noncommutative
geometry can be stated as follows:

\begin{itemize}

\item 

 I. Noncommutative special relativity gives the new
formulation of relativistic quantum mechanics which does not refer to a classical spacetime.
The standard quantum commutation relations of quantum mechanics are deduced from the spacetime and the momentum commutation relations. 

\item II. Noncommutative general relativity is quantum gravity. 

\end{itemize}

In the next sub-section we suggest the reformulation of quantum mechanics in terms of noncommuting coordinate systems, and the recovery of standard quantum mechanics from this reformulation. We then show how inclusion of self-gravity leads to a non-linear Schr\"{o}dinger equation. Much work still remains to be done, in terms of making rigorous contact with noncommutative differential geometry, and arriving at the field equations for noncommutative general relativity. However the flow of ideas appears rather natural, and we arrive at an explanation for collapse-induced quantum measurement which can be subjected to an experimental test. We would like to emphasize once again that the ideas in this section were not developed with the \textit{ad hoc} purpose of explaining quantum measurement, but have been concerned
with an altogether different aspect of quantum mechanics - the unsatisfactory presence of an external
classical time in the theory. 

\subsection{Quantum Minkowski Spacetime}
Consider a system of quantum mechanical particles having a total mass-energy
much less than Planck mass $m_{Pl}$, and assume that no external classical 
spacetime manifold is available. Since Planck mass scales inversely with the
gravitational constant, we are justified here in neglecting the gravitational
field, and the resulting quantum spacetime produced by the system will be
called a `quantum Minkowski spacetime'. 

To describe the dynamics using
noncommutative geometry consider a particle with mass $m\ll m_{Pl}$    
in a 2-d noncommutative spacetime with coordinates ($\hat{x}, \hat{t})$.
On the quantum Minkowski spacetime we introduce the non-Hermitean flat metric 
\begin{eqnarray}
\label{ncfm}  
\hat{\eta}_{\mu\nu} = \left(\begin{array}{cc}
                      1 & 1 \\
                      -1 & -1 \end{array} \right)
\end{eqnarray}
and the corresponding noncommutative line-element
\begin{equation}
ds^{2}=\hat{\eta}_{\mu\nu}d\hat{x}^{\mu}d\hat{x}^{\nu}=
d\hat{t}^{2}-d\hat{x}^{2}
+d\hat{t}d\hat{x}-d\hat{x}d\hat{t}
\label{lin}
\end{equation}
which is invariant under a generalized Lorentz transformation.

Noncommutative dynamics is constructed by formally defining a velocity  $\hat{u}^{i}=
d\hat{x}^{i}/ds$,
which, from (\ref{lin}), satisfies the relation
\begin{equation}
1=\hat{\eta}_{\mu\nu}\frac{d\hat{x}^{\mu}}{ds}\frac{d\hat{x}^{\nu}}{ds}=
(\hat{u}^{t})^{2}-(\hat{u}^{x})^{2} + 
\hat{u}^{t}\hat{u}^{x} - \hat{u}^{x}\hat{u}^{t}.  
\label{vel}
\end{equation} 
We define a generalized momentum as 
$\hat{p}^{i}=m\hat{u}^{i}$, 
which hence satisfies 
\begin{equation}
\hat{p}^{\mu}\hat{p}_{\mu} = m^{2}.
\label{nchj}
\end{equation}
Here, $\hat{p}_{\mu}=\hat{\eta}_{\mu\nu}\hat{p}^{\mu}$ is well-defined. Written
explicitly, this equation becomes
\begin{equation}
(\hat{p}^{t})^{2}-(\hat{p}^{x})^{2} + 
\hat{p}^{t}\hat{p}^{x} - \hat{p}^{x}\hat{p}^{t}  = m^{2}.
\label{nce}
\end{equation}

Dynamics is constructed by introducing a {\it complex} action 
$S(\hat{x},\hat{t})$ and by defining the momenta introduced above as
gradients of this complex action. In analogy with classical mechanics this
converts (\ref{nce}) into a (noncommutative) Hamilton-Jacobi equation,
which describes the dynamics.

When an external classical Universe with a classical manifold
$(x,t)$ becomes available (see below), one defines the generalized momentum 
$(p^{t},p^{x})$ in terms of the complex action $S(x,t)$ as 
\begin{equation}
p^{t}=-{\partial S\over \partial t}, \qquad p^{x} = 
{\partial S \over \partial x}
\label{pmoo}
\end{equation}
and from (\ref{nce}) the following fundamental rule for relating 
noncommutative dynamics to standard quantum dynamics 
\begin{equation}
(\hat{p}^{t})^{2}-(\hat{p}^{x})^{2} + 
\hat{p}^{t}\hat{p}^{x} - \hat{p}^{x}\hat{p}^{t}  = ({p}^{t})^{2}-({p}^{x})^{2}
 + i\hbar {\partial p^{\mu}\over \partial x^{\mu}}.
\label{nceq}
\end{equation}
A detailed justification for this key equation has been given in \cite{singh}.

In terms of the complex action the right hand side of this equation can be
written as
\begin{equation}
\left({\partial{S}\over \partial t}\right)^{2}-\left({\partial{S}\over \partial x}\right)^{2}
-i\hbar\left({\partial^{2}S\over\partial t^{2}}-{\partial^{2}S\over\partial x^{2}}\right)=m^2\label{hjc}
\end{equation}
and from here, by defining a quantum state $\psi$ in a natural manner:
 $\psi=e^{iS/\hbar}$, we arrive at the Klein-Gordon equation 
 \begin{equation}-\hbar^{2}
\left({\partial^{2}\over\partial t^{2}}-{\partial^{2}\over\partial x^{2}}\right)\psi=m^{2}\psi\label{kg}.
\end{equation}
In this manner we have arrived at standard quantum mechanics, starting from a formulation which did not
make reference to a classical time.  

The proposed commutation relations on the non-commutative spacetime are 
\begin{equation}
[\hat{t},\hat{x}]=iL_{Pl}^{2}, \qquad [\hat{p}^{t}, \hat{p}^{x}]=iP_{Pl}^{2}.
\label{commu}
\end{equation}
In \cite{singh} it has been suggested as to how one could infer the standard commutation relation of
quantum mechanics from the relations given above.

We have restricted the discussion here to a single particle in two dimensions.
The generalization to the multi-particle case, in four dimensions, is straightforward,
and outlined in \cite{singh}.

\subsection{Including Self-Gravity}
If the mass-energy of the particle is not negligible
in comparison to Planck mass its self-gravity must be taken into 
account. The `flat' metric (\ref{ncfm}) gets modified to the `curved' metric
\begin{eqnarray}
\label{nccm}  
\hat{h}_{\mu\nu} = \left(\begin{array}{cc}
                      \hat{g}_{tt} & \hat{\theta} \\
                      -\hat{\theta} & -\hat{g}_{xx} \end{array} \right)
\end{eqnarray}
We have made the significant assumption that in addition to the standard symmetric metric $\hat{g}_{\mu\nu}$
the noncommutative `curved' metric also has an antisymmetric component $\hat{\theta}_{\mu\nu}$. The
component $\hat{\theta}_{\mu\nu}$ will play a central role in our explanation of quantum measurement. 

With the introduction of the curved metric, Eqns.(\ref{lin}), (\ref{nchj}), (\ref{nceq}) and  
(\ref{hjc}) are respectively replaced by the plausible equations
\begin{equation}
ds^{2}=\hat{h}_{\mu\nu}d\hat{x}^{\mu}d\hat{x}^{\nu}=
\hat{g}_{tt}d\hat{t}^{2}-\hat{g}_{xx}d\hat{x}^{2}
+\hat\theta[d\hat{t}d\hat{x}-d\hat{x}d\hat{t}],
\label{linc}
\end{equation}
\begin{equation}
\hat{h}_{\mu\nu}\hat{p}^{\mu}\hat{p}^{\nu}=m^{2}, 
\label{casnc}
\end{equation}
\begin{equation}
\label{nceq2}\hat g_{tt}(\hat p^t)^2-\hat g_{xx}(\hat p^x)^2+\hat \theta
\left( \hat p^t\hat p^x-\hat p^x\hat p^t\right) =m^2,
\end{equation}
\begin{equation}
\label{nceq3}g_{tt}({p}^t)^2-g_{xx}({p}%
^x)^2+i\hbar \theta {\frac{\partial p^\mu }{\partial x^\mu }}=m^2.
\end{equation}

These replacements have been made very much in the same spirit in which one goes from
flat spacetime equations to curved spacetime equations in classical general relativity.
Like in general relativity, the metric $\hat{h}_{\mu\nu}$ is assumed to be
determined by the mass $m$ via the quantum state $S(\hat{x},\hat{t})$. 
The field equations are assumed to be covariant under general coordinate transformations
of noncommuting coordinates. If these field equations could be determined, they would
constitute the field equations of quantum gravity, in this approach.

In the macroscopic limit 
$m\gg m_{Pl}$ the antisymmetric component $\theta$ is assumed to go to zero; the 
noncommutative spacetime (\ref{linc}) is then indistinguishable from ordinary 
commutative spacetime, and Eqn. (\ref{nceq3}) reduces to classical dynamics. 
In the microscopic limit $m\ll m_{Pl}$ we have that $\theta$ goes to one,
$g_{tt}$ and $g_{xx}$ also go to one,
and we recover standard quantum mechanics. Thus we see that when $\theta$ is different from zero and one, we get a new mechanics which is neither standard linear quantum mechanics, nor classical mechanics!

There are, however, issues which remain to be resolved. We have assumed $\theta$ to be real, which
appears a reasonable choice, considering that it represents an additional component of the gravitational
field, significant only in the mesoscopic domain. We do not know at the moment the explicit spacetime dependence of $\theta$ - we will assume as of now that $\theta$ depends only on the mass $m$, and on the quantum state $S(x,t)$.

\subsection{A non-linear Schr\"{o}dinger equation}
If we substitute
for the momenta in (\ref{nceq3}) in terms of the complex action using
(\ref{pmoo}) and then substitute $\psi=e^{iS/\hbar}$ and take the
non-relativistic limit, the resulting effective Schr\"{o}dinger equation is
non-linear \cite{singh}. It is very similar to the Doebner-Goldin equation discussed before - the latter arises when one classifies physically different quantum systems by considering unitary representations of the group of diffeomorphisms $Diff(R^{3})$. 

The simplest case is obtained when in 
(\ref{nceq3}) one approximates the diagonal metric components to unity,
giving the non-linear Schr\"{o}dinger equation 
\begin{equation}
i\hbar\frac{\partial\psi}{\partial t} = -\frac{\hbar^{2}}{2m}\frac
{\partial^{2}\psi}{\partial x^{2}} + \frac{\hbar^{2}}{2m}(1-\theta)
\left(\frac{\partial^{2}\psi}{\partial x^{2}} - [(\ln\psi)']^{2}\psi
\right) + V(x)\psi.
\label{nlse}
\end{equation}
We have generalized by including a potential $V(x)$. This equation bears a striking resemblance to
the Doebner-Goldin equation (\ref{dg}); the two equations have been compared in detail in \cite{singh}.

This equation can be rewritten as,
\begin{equation}
i\hbar\frac{\partial\psi}{\partial t} = -\frac{\hbar^{2}}{2m}\frac
{\partial^{2}\psi}{\partial x^{2}} + \frac{\hbar^{2}}{2m}(1-\theta)
\left(\frac{\partial^{2}[\ln\psi]}{\partial x^{2}}
\right)\psi + V(x)\psi,
\label{nlses}
\end{equation}
and then more usefully, by expanding the non-linear term into real and imaginary parts, as
\begin{equation}
i\hbar\frac{\partial\psi}{\partial t} = -\frac{\hbar^{2}}{2m}
\frac{\partial^{2}\psi}{\partial x^{2}} + V(x)\psi + 
\frac{\gamma(m)\hbar^{2}}{2m}q\frac{\partial^{2}(\ln R)}{\partial x^{2}}\psi + 
i\frac{\gamma(m)\hbar}{2m}q\frac{\partial^{2}\phi}{\partial x^{2}}\psi + V(x)\psi
\label{goe}
\end{equation}
where $\gamma(m)q=(1-\theta)$ and $\psi=Re^{i\phi/\hbar}$. We have made the plausible assumption that
$(1-\theta)$ can be written as a product of two positive terms - a part $\gamma(m)$ which does not depend on the state, and a part $q$ which depends on the state, but not on the mass.
We see in the second-last term of (\ref{goe}) the emergence of the non-Hermitean, non-linear part which is of interest to us. Gravity is responsible for this term because $\theta$ is actually a function
of $m/m_{Pl}$, and so is $\gamma$.

It may appear that the non-linear equation we have derived is complicated, and does not possess the simplicity of the linear Schr\"{o}dinger equation. However it is worth recalling that this equation is
the non-relativistic limit of the highly symmetric Eqn. (\ref{casnc}). The relativistic equation, when written in noncommuting coordinates, and in terms of the complex action, has a symmetric and simple form.

The above equation is thus similar to the non-linear Schr\"{o}dinger equation (\ref{gri}) reviewed
in the previous section. However, our equation is not norm-preserving! It is norm-preserving if the
probability density is defined as $|\psi|^{2/\theta}$, instead of $|\psi|^2$. We need not regard 
this circumstance as an implausible one, since in this mesoscopic domain (where
$\theta$ is neither one nor zero) we would not know a priori what the exact definition of norm, in terms of the wave-function, is \cite{singh}.

One could ask for a reason for the presence of the non-Hermitean term in the Hamiltonian in Eqn. (\ref{nlse}). The answer is that the presence of such a term is generic; it is only in the small
mass, linear, limit that this term is negligible. Further, so long as the evolution is norm-preserving, the 
presence of such a term cannot be regarded as objectionable.  

In passing, we note that 
in terms of the complex action function $S$ defined earlier as $\psi=e^{iS/\hbar}$ the non-linear Schr\"{o}dinger equation (\ref{nlse}) is written as
\begin{equation}
\frac{\partial S}{\partial t} = - \frac{S'^{2}}{2m} + \frac{i\hbar}{2m}\theta(m)S''\ + V(x) .
\label{nrhj}
\end{equation}
This equation is to be regarded as the non-relativistic limit of Eqn. (\ref{nceq3}). We easily see here
that in the limit $\theta=0$ classical mechanics is recovered; and that setting $\theta=1$ gives
the linear Schr\"{o}dinger equation. The intermediate regime, where $\theta$ is neither one nor zero, is different from both classical and quantum mechanics. This regime is consistent both with classical and
quantum mechanics, but will go undetected in experiments unless one examines properties of mesoscopic systems. Interestingly enough, if $\theta$ is different from one, the non-linear equation can explain the collapse of the wave-function, as we will now see.

\subsection{Explaining Quantum Measurement}
Prior to the onset of a quantum measurement, evolution is described by
\begin{equation}
i\hbar\frac{\partial\psi}{\partial t} = -\frac{\hbar^{2}}{2m}
\frac{\partial^{2}\psi}{\partial x^{2}} + V(x)\psi
\end{equation}
thus preserving superposition. This is because we have $m\ll m_{Pl}$ and $\theta\rightarrow 1$.

We recall that the onset of 
measurement corresponds to mapping the state $|\psi>$ to 
the state $|\psi>_F$ of the final system as
\begin{equation}
|\psi>\rightarrow |\psi>_F\ \equiv  \sum_n a_n|\psi>_{Fn}= \sum_n\ a_n|\psi_n>|A_n>
\label{map}
\end{equation}
where $|A_n>$ is the state the measuring apparatus would result in, had the 
initial  system been in the state $|\psi_n>$.

At the onset of measurement, evolution is described by the equation
\begin{equation}
i\hbar\frac{\partial\psi_F}{\partial t} = H_F \psi_F +
\frac{\gamma(m)\hbar^{2}}{2m_F}q\frac{\partial^{2}(\ln R_F)}{\partial x^{2}}\psi_F + 
i\frac{\gamma(m)\hbar}{2m_F}q\frac{\partial^{2}\phi_F}{\partial x^{2}}\psi_{F}.
\label{nll}
\end{equation}
$H_F$ is the Hermitean part of the Hamiltonian for the final system (including both quantum
system and measuring apparatus).
Also, $\gamma(m_F)q=(1-\theta_F)$ and $\psi_F=R_Fe^{i\phi_F/\hbar}$.
This equation should be compared with the non-linear equation in the toy model.

The states $|\psi>_{Fn}$ cannot evolve as a superposition because the
evolution is now non-linear.
However, the initial state at the onset of measurement \textit{is} a superposition of the $|\psi_{Fn}>$;
it is simply the entangled state $|\psi>_F$ at the onset of measurement.
This initial superposition must thus break down during further evolution, according to the law
\begin{equation}
i\hbar\frac{\partial a_{n}}{\partial t} =  
i\frac{\gamma(m_F)\hbar}{2m_F}q\frac{\partial^{2}\phi_F}{\partial x^{2}}a_{n}
\label{supbreak}
\end{equation}
which follows after substituting the expansion (\ref{map}) in (\ref{nll}).
Here, we ignore the Hermitean part of the Hamiltonian, and focus only on the decay/growth of the quantum state. Note that the $q_n$'s are different for different states. This is because it is natural to assume that the component $\theta$ of the gravitational field, which is produced by the mass $m$, should depend on the quantum state. Also, $\phi_F$ is the value of the phase of the state $|\psi>_F$ at the onset of measurement.

We thus get, as before 
\begin{equation}
\frac{d}{dt}\ln\frac{a_i}{a_j} = \frac{\gamma}{2m_F}(q_i-q_j)\phi_F''
\end{equation}
and like for the toy model, only the state with the largest $q$ survives. This is ensured because
the $a_n$ satisfy the condition $\Sigma |a_n|^2=1$ because of the initial conditions imposed on them.
In addition, as noted above, $|\psi|^{2/\theta_F}$, and hence $(\Sigma|a_n|^2)^{1/\theta_F}$ is preserved
during evolution. Its important to note the subtlety that there will be a $\theta_F$ associated
with the state $|\psi>_F$ and different $\theta_{Fn}$ associated with each of the states $|\psi>_{Fn}$. 

In order to recover the
Born rule, the $q_n$ must be random variables.
Only further development in theory can determine if this is so, and what their probability distribution is. One could nonetheless, following Grigorenko, assign a probability distribution for the 
$q_n$ so as to recover the Born rule. Once again, the phase of the initial quantum state $\psi(t_0)$
is an attractive candidate for the desired random variables.

From Eqn. (\ref{supbreak}) we can define an important quantity, the lifetime 
$\tau_{sup}$ of a superposition. It can be read off from this equation to be
\begin{equation}
\tau_{sup} = \frac{2m}{(1-\theta)\phi_F''}. 
\label{sup1}
\end{equation}
The first inference we can draw is that, since $\theta$ is strictly equal to one in standard linear quantum mechanics, a quantum superposition has an infinite lifetime in the linear theory, as one would expect. However, the situation begins to change in an interesting manner as the value of the mass
$m$ approaches and exceeds $m_{Pl}$. Since we know that in this limit $\theta$ approaches zero, we can
neglect $\theta$, and the superposition lifetime will then essentially be given by
\begin{equation}
\tau_{sup} \approx \frac{m}{\phi_F''} \sim \frac{mL^2}{\phi} 
\label{sup2}
\end{equation}
where $L$ is the linear dimension of the system, and $\phi_F$ is the phase of the state
$|\psi>_F$ at the onset of measurement. 
For a macroscopic system we can get a numerical estimate of the life-time by noting that we are close to the classical limit, where the phase coincides with
the classical action in the Hamilton-Jacobi equation. To leading order, the magnitude of the classical action is given by $S_{cl}=mc^{2}t$, where $t$ is the time over which we observe the classical trajectory; approximately, this could be taken to be the value of the phase $\phi$, and $\tau_{sup}$ is
then roughly given by
\begin{equation}
\tau_{sup}\sim \frac{1}{t}\left(\frac{L}{c}\right)^{2}.
\label{sup3}
\end{equation}
For a measuring apparatus, if we take the linear dimension to be say $1$ cm, and the time of observation to be say $10^{-3}$ seconds, we get the superposition lifetime to be $10^{-18}$ seconds, which 
is an encouragingly small number. This could possibly explain why the wave-function collapses so rapidly during a measurement.

We can get a very rough estimate of $\tau_{sup}$ for a mesoscopic system using
(\ref{sup3}), and continuing to ignore $\theta$. Let us take $L\sim 10^{-3}$ cm and
correspondingly  $m \sim 10^{-9}$ gm. Such a composite object has approximately $10^{15}$ particles, and we could take  $\phi\sim N\hbar$ with $N\sim 10^{15}$. This gives $\tau_{sup}\sim 10^{-3}$ seconds. 

Thus, in making a transition from a microscopic system which obeys linear quantum mechanics, to
a macroscopic system such as a measuring apparatus, we find that the lifetime of a superposition
changes from an astronomically large value to an immeasurably small value. We could thus be certain
that there must exist intermediate, mesoscopic systems for which the lifetime of a superposition is
an easily measurable number, say one milli-second. Unfortunately our present understanding of the approach described here is not good enough to say at what value of the mass $m$ this will happen. What we can be certain about is that if the ideas described here are on the right track, then as experimentalists 
check for quantum superposition in larger and larger quantum systems, they will discover systems for which the lifetime of a superposition will become small enough to be measurable in the laboratory
in principle. In practice, it remains to be seen whether or not decoherence will permit this
lifetime to be measured.

\subsection{Ideas for an Experimental Test of the Model}
The possibilities for an experimental test of this non-linear model, and of its implications
for quantum measurement, can be divided into three classes:

(i) Looking for breakdown of quantum superposition: This suggestion is in line with the kind of experiments already in progress - constructing larger and larger composite quantum objects 
(i.e. Carbon-60 and beyond) and checking whether or not linear superposition holds for such objects.
Our prediction is that by the time the number of particles in the composite object reaches to about
$10^{15}$ particles, the lifetime of the superposition will become small enough to be observable in the
laboratory, and one will actually observe the decay of superposition in such a system.
The greatest obstacle to the detection of such an effect, even if it is there, will come from
the phenomenon of decoherence. In order to ascertain whether or not there is a breakdown of non-linearity
due to superposition, one will have to first ensure that the object is sufficiently well-isolated from its environment, so that decoherence can be avoided.

(ii) Difference between the predictions of the linear theory and the non-linear theory: If one were to
calculate expectation values of observables, using the non-linear equation (\ref{nlse}), the result
will be different from what one will get from the linear theory, because now $\theta$ is non-zero.
This aspect has been discussed in some detail in \cite{singh}. Care has to be taken to construct gauge-invariant observables for the non-linear theory. It was shown in \cite{singh} that the ratio $\hbar/m$ is one such gauge-invariant observable - the non-linear model predicts that the effective Planck's constant in the theory is $\hbar\theta(m)/m$. Thus, according to the non-linear model, an experiment to measure the value of $\hbar/m$ for a mesoscopic system will give results consistent with theory only if one assumes Planck's constant to have an effective value $\hbar\theta(m)$.

(iii) Looking for a correlation between the absolute value of the initial phase, and the outcome of a quantum measurement : If the random variable responsible for the outcome of a quantum measurement in the non-linear Schr\"{o}dinger equation (\ref{nll}) is indeed the absolute phase of the initial state $\psi(t_0)$, then the correlation between this phase and the outcome of a measurement should in principle be detectable by experiment. Conventional wisdom is that absolute phase cannot be measured and that only
differences in phase are measurable. However, if the absolute value of the phase is going to decide the outcome of a measurement, it does not seem unnatural to expect that there might be a way to determine the
value of this initial phase. In other words, while absolute phase is not observable in the linear theory,
it probably does become observable in the non-linear theory, and one should explore possible ways to
measure it in the non-linear theory, i.e. for mesoscopic systems.

\section{Other Models for Collapse of the Wave-function}
Various researchers have proposed modifications of the Schr\"{o}dinger equation, with the purpose
of explaining quantum measurement as a dynamically induced collapse. Most of these models also 
estimate the lifetime of a quantum superposition - the lifetime being large for microscopic systems, and small for macroscopic ones. While some models do not involve gravity, it is
remarkable that many of the models assert that gravity is responsible for collapse. Below,
we review some of the models very briefly, without really attempting to critically compare these models with each other, or with our own approach. (A comparison across various models of collapse will be reported elsewhere.  The literature on the subject is large, and we confine ourselves to giving pointers to the literature. A nice, though somewhat older, review of collapse models has been given by 
Pearle \cite{Pearle2}).
\subsection{Models that do not Involve Gravity}
{\bf Dynamical Reduction Models based on Non-linear Stochasticity} : The first suggestion that collapse
could be explained as a dynamical reduction process using stochastic differential equations came from Pearle \cite{Pearle3}. The Schr\"{o}dinger equation was to be augmented by a stochastic term which could induce collapse. Significant development in this direction came from the work of Ghirardi, Rimini and Weber (GRW) \cite{GRW}. This program, and its progress,  has been reviewed in \cite{BG} and in 
\cite{Bassi}. There were two guiding principles for this dynamical reduction model (known as QMSL: Quantum Mechanics with Spontaneous Localization) \cite{BG} :

"
1. The preferred `basis' - the basis on which reductions take place - must be chosen in such a way as to guarantee a definite position in space to macroscopic objects.

2. The modified dynamics must have little impact on microscopic objects, but at the same time must reduce the superposition of different macroscopic states of macro-systems. There must then be an amplification mechanism when moving from the micro to the macro level.  "

The reduction is achieved by making the following set of assumptions:

"
1. Each particle of a system of $n$ distinguishable particles experiences, with a mean rate $\lambda_i$,
a sudden spontaneous localization process.

2. In the time interval between two successive spontaneous processes the system evolves according to the usual Schr\"{o}dinger equation. "

In their model, GRW introduced two new fundamental constants of nature, assumed to have definite numerical values, so as to reproduce observed features of the microscopic and macroscopic world.
The first constant, $\lambda^{-1}\sim 10^{16}$ seconds, alluded to above, determines the rate of spontaneous localization (collapse) for a single particle. For a composite object of $n$ particles,
the collapse rate is $(\lambda n)^{-1}$ seconds. The second fundamental constant is a length scale 
$a\sim 10^{-5}$ cm which is related to the concept that a widely spaced wave-function collapses to a length scale of about $a$ during the localization. 

A gravity based implementation of the QMSL model has been studied by Diosi \cite{Diosi} and generalized in \cite{g}. 

The QMSL model has the limitation that it does not preserve symmetry of the wave-function under
particle exchange, and has been improved into what is known as the CSL (Continuous Spontaneous Localization) model \cite{p4}, \cite{gpr}. In CSL a randomly fluctuating classical field couples
with the particle number density operator of a quantum system to produce collapse towards its
spatially localized eigenstates. The narrowing of the wavefunction amounts to an increase in the particle's energy, and actually amounts to a violation of energy conservation. This intriguing aspect of
the collapse model has been discussed in \cite{p4} and an interesting suggestion for preserving energy conservation has been made therein. 

An outstanding open question with regard to the dynamical reduction models is the origin of the random noise, or the randomly fluctuating classical scalar field, which induces collapse. We see herein the
possibility of a connection, worth exploring further, with our proposal that the non-linear Schr\"{o}dinger equation has its origin in quantum gravity. Could the randomly fluctuating classical field of CSL 
be related to the random parameter $\theta(m)$ we have in our non-linear equation?

Another collapse model is due to Adler (see \cite{Adler} and related references therein), where an energy-driven stochastic Schr\"{o}dinger equation is a phenomenological model for state-vector reduction.
Collapse in this model is energy conserving and reproduces the Born probability rule. It is interesting
to note that in this model, the terms that influence wave-vector reduction are directly related to
mass-energy, and hence once again a connection with gravity is suggested.

\subsection{Models that Involve Gravity} 
The gravitational field produced by a classical object obeys the laws of general relativity, or in the limiting case, those of Newtonian gravity. Since the position of the classical object is subject
to tiny quantum uncertainties, the gravitational field and the curvature tensor produced by it are also subject to quantum fluctuations. Karolyhazy \cite{Karolyhazy}, \cite{k2} built an interesting and
plausible model to explain how quantum superposition could be destroyed in macroscopic objects, as a result of these quantum fluctuations in the gravitational field of the object. The possible connection
of such a model with the CSL approach has been discussed in \cite{Pearle2}. Other models discussing the
possible role of gravity in collapse are presented in \cite{gr1}, \cite{gr2}.

In spirit and concept, our work here comes closest to Penrose's idea that gravity is responsible for
wave-vector reduction \cite{pen}. Penrose argues convincingly that the principle of general covariance
and the principle of linear superposition in quantum mechanics are in direct conflict with each other.
Our starting point has been essentially the same - we argued that because of general covariance, one
cannot have a classical spacetime manifold coexisting with a Universe which has only quantum matter fields. This lead us to conclude that if gravity cannot be neglected, quantum theory must be 
non-linear - which is essentially what Penrose has argued: general covariance and linear superposition
are incompatible  with each other.

Penrose develops an estimate for the lifetime of the quantum superposition of two different
positions of a macroscopic object, and demonstrates it to be of the order $\hbar/\Delta E$, where 
$\Delta E$ is the gravitational self-energy of the difference between the mass distributions of
each of the two locations of the macroscopic object. An experimental test of Penrose's idea has been
proposed in \cite{pex}.

\section{Discussion}

In putting forth the two possible explanations of quantum measurement, we have assumed that the wave-function describes an individual quantum system, and not a statistical ensemble of quantum systems.
Also, we have ignored making any mention of the Copenhagen interpretation, which essentially states
that upon measurement, the wave-function collapses into one of the eigenstates, but the interpretation does not suggest any mechanism for the collapse.

We believe that our case for the necessity of a reformulation of quantum mechanics is robust. Equally robust is the inference that the standard linear quantum theory is a limiting case of a non-linear quantum theory. However, only partial justification can be given for the use of noncommutative differential geometry for constructing such a reformulation. Having accepted to work in the framework 
of noncommutative geometry, we regard it as highly attractive that linear quantum theory, and its 
nonlinear generalization, are respectively the equations of motion in a noncommutative special 
relativity, and in its generalization to a noncommutative general relativity. Various issues here remain to be understood much better. These include: (i) the physical meaning associated with the noncommutative metric (\ref{ncfm}), (ii) the nature of commutation relations to be imposed on the `curved' noncommutative metric $\hat{h}_{\mu\nu}$, (iii) the full development of the concept of general covariance in noncommutative geometry,  and the related generalization of the concept of curvature (for a review of the current status of this aspect see for instance \cite{connes}, \cite{Madore}),(iv) the field equations, analogous to those in general relativity, for this metric,  (v) the justification for retaining $\theta(m)$ in the description of the nonlinear Schr\"{o}dinger equation (\ref{nlse}), while ignoring the diagonal components $g_{\mu\nu}$.  

In spite of these important issues which are yet to be addressed and resolved, we regard it as a natural consequence of the required reformulation that there is a new mechanics in the intermediate, mesoscopic domain; and that classical mechanics, as well as linear quantum mechanics, are its limiting cases. The
limits are obtained by letting $\theta\rightarrow 0$ and $\theta\rightarrow 1$ respectively. 

\textit{We re-emphasize that there is no reason to believe a priori that quantum mechanics continues to hold
unchanged in the mesoscopic domain. And if there indeed are reasons to expect a departure from standard quantum theory in this domain (and these reasons are independent of the issue of quantum measurement) then experiments must be carried out to test the laboratory predictions of these new ideas. 
}

There are very stringent experimental bounds on the presence of non-linear terms in quantum mechanics in
the atomic domain \cite{weinberg}, \cite{weinberg2}, \cite{expt}. However, these bounds do not extend to the mesoscopic
domain - thus for instance there are no bounds on non-linear quantum mechanics when such a theory is applied to an object containing say $10^{15}$ particles. 

{\bf Superluminality:} It has also been pointed out that the presence of a non-linearity in quantum mechanics can result in
the possibility of superluminal communication \cite{gi}, \cite{polchinski}. In our approach, the non-linearity
is a relic of a more fundamental description of the theory in terms of a noncommutative spacetime.
We do not have at present a good understanding of the `light-cone structure' in a noncommutative spacetime, and it is difficult to assess the possibility or otherwise of superluminal communication in a noncommutative geometry. More importantly, the non-linearity we predict becomes significant
only in the mesoscopic domain, and we are suggesting that mechanical laws here are different 
from both the classical and the quantum case. Thus in this domain the issue of superluminality
needs to be addressed afresh. Once again, it needs to be stressed that the effects of non-linearity
could be severely masked by decoherence resulting from interaction with the environment, making the
mesoscopic mechanics effectively indistinguishable from classical or quantum mechanics.

With regard to superluminality, it has also been pointed out by Doebner and Goldin \cite{goldin2} that
by virtue of non-linear gauge transformations, many non-linear Schr\"{o}dinger equations are
physically equivalent to linear equations. So one could not possibly deduce superluminal communication for such non-linear equations.

\textbf{The preferred basis problem:} We would like to suggest, following GRW, that a preferred basis is one
in which positions of macroscopic objects are localized. This is consistent with the oft
expressed view that "\textit{all} quantum mechanical measurements consist of or are obtained from positional measurements made at various times \cite{goldin2}".  As stated by Feynman and Hibbs \cite{fh}, p. 96

"Indeed all measurements of quantum mechanical systems could be made to reduce eventually to position
and time measurements. Because of this possibility a theory formulated in terms of position
measurements is complete enough in principle to describe all phenomena".

\bigskip

\noindent
{\bf Acknowledgements:} For useful discussions and conversations during various stages of this work, it is a pleasure to thank  Joy Christian, Patrick Das Gupta, Atri Deshamukhya, Avinash Dhar, 
Hanz-Dietrich Doebner, Thomas Filk, Gerald Goldin,
T. R. Govindarajan, Kumar S. Gupta, Sashideep Gutti, Friedrich Hehl, Eric Joos, F. Karolyhazy, Romesh Kaul,  Claus Kiefer, Gautam Mandal, Shiraz Minwalla, Ayan Mukhopadhyay, T. Padmanabhan, Aseem Paranjape,
Rainer Plaga, L. Sriramkumar, Rakesh Tibrewala, A. V. Toporensky, Sandip Trivedi,  C. S. Unnikrishnan, Cenalo Vaz, Albrecht von M\"{u}ller and Spenta Wadia.

\end{document}